# Energy Efficient Multi-Level Clustering To Prolong The Lifetime of Wireless Sensor Networks

Surender Soni and Narottam Chand

**Abstract**— Clustering in wireless sensor networks (WSNs) is an important technique to ease topology management and routing. Clustering provides an effective method for prolonging lifetime of a WSN. This paper proposes energy efficient multi-level clustering schemes for wireless sensor networks. Wireless sensor nodes are extremely energy constrained with a limited transmission range. Due to large area of deployment, the network needs to have a multi-level clustering protocol that will enable far-off nodes to communicate with the base station. Simulation is used to analyze the proposed protocols and compare their performance with existing protocol EEMC [5]. Simulation results demonstrate that our proposed protocols are effective in prolonging the network lifetime.

## 1 Introduction

RECENT technological advancements have enabled the deployment of large scale wireless sensor networks for real-world applications. Inexpensive sensors are deployed for data collection from the field in a variety of scenarios including health, military surveillance, building security, in harsh physical environments, for scientific investigations on other planets, etc. [1, 2]. A sensor node has limited computing capability and memory, and it operates with limited battery power. These sensor nodes can self organize to form a network and can communicate with each other in a wireless manner [1, 3].

Each node has transmitter power control and an omni-directional antenna, and therefore can adjust the area of coverage with its wireless transmission. Typically, sensor nodes collect audio, seismic, and other types of data and collaborate to perform a high level task in a sensor web. For example, a sensor network can be used for detecting the presence of potential threats in a military conflict. Since wireless communications consume significant amounts of battery power, sensor nodes should be energy efficient in transmitting data [4, 6, 7].

WSNs are usually employed to cover a very large area and thus need to be in touch with far away base station. Clustering is an effective topology control approach in WSNs which can increase network scalability and lifetime. In real-world scenarios all the nodes may not be able to reach the base station directly. This leads to need for multi-level clustering algorithms that will form multi-level cluster heads (CHs). Loosely speaking, a multi-level clustering algorithm is one in which nodes are arranged in a hierarchical cluster setup such that Level-i CH sends aggregated data to Level-(i-1) CH which in turn reports data to Level-(i-2) CH and so on until the Level-1 CH receives data and forwards it to base station.

In [5], the authors propose an energy-efficient multi-level clustering algorithm called Multi-Level Clustering Algorithm (EEMC), which aims at minimum energy consumption in sensor networks. EEMC also covers the cluster head election scheme. In EEMC, the data collection operation is broken up into rounds, where each round begins with a cluster set-up phase, which means that the nodes execute EEMC algorithm to form a multi-level clustering topology independently, and continues with a data transmission phase, which means the nodes transmit the sensed data packets to the sink node under such a clustering topology. Assuming that sink node is remotely located and sensor nodes are stationary, simulation results show that their proposed algorithm is effective in prolonging the network lifetime of a large-scale network. They also show that the algorithm has low latency and moderate overhead across the network [5].

The EEMC [5] algorithm has the limitation that the regular nodes can join the last level of CHs only, thus incurring high latency in the network. Another notable limitation is that each node be GPS equipped to know its location precisely. If the precise location is not known, the algorithm will fail. In order to overcome these shortcomings, we propose two new algorithms, LAMC (Location Aware Multi-level Clustering) and PAMC (Power Aware Multi-level Clustering). Simulations are used to analyze the performance of proposed algorithms.

The remainder of the paper is organized as follows. Section 2 describes the assumptions and the energy model used in this paper, Section 3 outlines the EEMC algorithm

- *Surender Soni, Department of Computer Science & Engineering, National Institute of Technoloyg Hamirpur, India.*
- *Narottam Chand, Department of Computer Science & Engineering, National Institute of Technoloyg Hamirpur, India.*

proposed in [5] and in Section 4 we propose two new algorithms for energy efficient multi-level clustering. Section 5 discusses the simulation results of the proposed algorithms and compares it with the existing ones. Concluding remarks are given in Section 6.

## 2 PRELIMINARIES

In this paper, we consider a sensor network consisting of N nodes randomly deployed over a vast field to continuously monitor the environment. We make some assumptions about the sensor nodes and the underlying network model:

- There is a base station (i.e. sink node) located far from the sensing field. Sensors and the base station are all stationary after deployment.
- All nodes in the network are homogenous and energy constrained.
- All nodes are able to send data to the base station.
- Propagation channel is assumed to be symmetric.
- Cluster-heads can aggregate data gathered into a single packet.

We use the same energy and radio model as in [5] with the value of electronics energy, $E_{elec} = 50nJ/bit$ as the energy being used to run transmitter and receiver circuits. The energy used for transmission amplifier $\varepsilon_{amp} = 10\frac{pJ}{bit}/m^2$. The energy cost of transmission ($E_{Tx}$) and reception ($E_{Rx}$) are calculated as

$$E_{Tx}(k,d) = E_{elec}k + \varepsilon_{amp}kd^2$$

$$E_{Rx}(k,d) = E_{elec}k$$

Where k is the length of transmitted/received message in bits and d is the Euclidian distance between sending and receiving nodes. Additionally energy is also used for aggregation of data at the cluster heads [8].

## 3 OVERVIEW OF EEMC ALGORITHM

In addition to the above assumptions, EEMC [5] also assumes that the nodes have location information. As detailed in [5], EEMC algorithm is divided into two phases:

**(a) Cluster Setup Phase**

(i) At the beginning of each round, all nodes are set to regular (non-CH status). The cluster set-up phase is initiated by each active node u sending its location information and its current residual energy $E_u(t)$ to the sink node to indicate that level-1 CH will now be selected.

(i) (ii) After the sink node receives these values, it will broadcast a ''command'' message with the following two values: one is the total remaining energy of the network $\sum E_v(t)$ and the other is $\sum 1/dis(n_v, n_{sink})$ the total reciprocal of distance from all active nodes. Here, $dis(n_v, n_{sink})$ is the distance between active node v and the sink node.

(iii) On receiving this ''command'' message, each active node u sets its probability of becoming a level-1 CH, denoted as $p_1(u)$, by using the following formula,

$$p_1(u) = N_{CH1}^{opt}\left[\phi\frac{E_u(t)}{\sum_{v\in S(t)}E_v(t)} + (1-\phi)\left(\frac{1/dis(n_u,n_{sink})}{\sum_{v\in S(t)}1/dis(n_v,n_{sink})}\right)\right]$$

Here, $\phi$ is a parameter and $N_{CH1}^{opt}$ is the optimal number of cluster heads for level-1. Thus, nodes with higher energy or less distance will have a greater probability of becoming the cluster heads.

(iv) After this, in order to reduce energy consumption, all level-1 CH send out a level-1 CH message with the range $\frac{R}{\sqrt{N_{CH1}^{opt}}}$. Any node receiving this message replies with a "join" message sending out its residual energy and the current position. After the level-1 CH (denoted as $n_{CH1}$) receive this message, they can construct their active node-set, denoted as $S_1(t)$, and then will broadcast a "command" message with total residual energy of the cluster and the total reciprocal distance within the cluster of each node from the cluster-head and also cardinality of $S_1(t)$ denoted as $N_1$. Each node will then set its probability of becoming a level-2 cluster-head according to the formula,

$$p_2(u) = \sqrt{N_1}\left[\phi\frac{E_u(t)}{\sum_{v\in S_1(t)}E_v(t)} + (1-\phi)\left(\frac{1/dis(n_u,n_{CH1})}{\sum_{v\in S_1(t)}1/dis(n_v,n_{CH1})}\right)\right]$$

Based on this probability, nodes elect whether they will become cluster-heads for level 2 or not.

(v) Generalizing, the clustering may extend up to j-levels. For any level-(j-1) CH, the CH would broadcast a level-(j-1) CH message with the radio range $\frac{R}{\sqrt{N_{CH1}^{opt}\prod_{j=1}^{j=2}\sqrt{N_j(t)}}}$.

Each regular node that receives this message will send its current position as well as the energy level to CH with a "join" message. After the level-(j-1) CHs (denoted as $n_{CHj-1}$) receive this message, they can construct their active node-set, denoted as $S_{j-1}(t)$, and then will broadcast a "command" message with total residual energy of the cluster and the total reciprocal distance within the cluster of each node from the cluster-head and also cardinality of $S_{j-1}(t)$ denoted as $N_{j-1}$. Each node will then set its probability of becoming a level-j cluster-head according to the formula,

$$p_j(u) = \sqrt{N_{j-1}}\left[\phi\frac{E_u(t)}{\sum_{v\in S_{j-1}(t)}E_v(t)} + (1-\phi)\left(\frac{1/dis(n_u,n_{CHj-1})}{\sum_{v\in S_{j-1}(t)}1/dis(n_v,n_{CHj-1})}\right)\right]$$

The process of Clustering will stop if a node has two or less number of nodes in its regular set.

**(b) Network Operation Phase**

After level-T clustering topology is formed, the regular nodes start transmitting the sensed data to lowest CHs. Level-T CHs aggregate the sensed data and send it to level-(T-1) CHs and so forth. Finally, all level-1 CHs transmit the aggregated data to sink node.

The cost of delivering data to the sink is sum of energy spent by nodes to send the data to their respective CH.

## 4 PROPOSED ALGORITHM

The EEMC algorithm discussed above has the following perceived shortcomings:

(i) The regular nodes can join the last level of CHs only. Thus the latency of the network is high since packets take a longer route to reach the base station.

(ii) The nodes need to be equipped with GPS systems in order to know their location precisely. If the precise location is not known, the algorithm will fail.

In order to overcome these shortcomings, we propose two new algorithms, LAMC (Location Aware Multi-level Clustering) and PAMC (Power Aware Multi-level Clustering). The basic idea of both these algorithms is borrowed from EEMC [5].

### 4.1 LAMC

The Location Aware Multi-level Clustering (LAMC) algorithm is same as EEMC, but without the shortcoming (i) listed above. The operation of LAMC is also performed in two steps.

**(a) Cluster Setup Phase**

(i) At the beginning of each round, all nodes are set to regular (non-CH status). The cluster set-up phase is initiated by the sink node sending a beacon signal which contains the current round number to each active node u present within its transmission range.

(ii) On receiving this beacon signal, each node sends its location information and its current residual energy $E_u(t)$ to the sink node to indicate that level-1 CH will now be selected.

(iii) After the sink node receives these values, it will broadcast a ''command'' message with the following two values: one is the total remaining energy of the network $\sum E_v(t)$ and the other is $\sum 1/dis(n_v, n_{\sin k})$ the total reciprocal of distance from all active nodes. Here, $dis(n_v, n_{\sin k})$ is the distance between active node v and sink node.

(iv) On receiving this ''command'' message, each active node u sets its probability of becoming a level-1 CH, denoted as $P_1(u)$, by using the following formula,

$$p_1(u) = N_{CH1}^{opt}\left[\phi\frac{E_u(t)}{\sum_{v\in S(t)}E_v(t)} + (1-\phi)\left(\frac{1/dis(n_u, n_{\sin k})}{\sum_{v\in S(t)}1/dis(n_v, n_{\sin k})}\right)\right]$$

Here, $\phi$ is a parameter which defines how much weightage is given to residual energy and how much to distance of node from sink and $N_{CH1}^{opt}$ is the optimal number of cluster heads for level-1 and is taken to be equal to $\sqrt{N}$ where N is the total number of nodes in the network. Thus, nodes with higher energy or less distance will have a greater probability of becoming the cluster heads. Nodes elect whether to become a level-1 cluster head or not based on global knowledge.

(v) After this, in order to reduce energy consumption, all level-1 CH send out a "level-1 CH" message with the range $\frac{R}{\sqrt{N_{CH1}^{opt}}}$. Any node receiving this message replies with a "join" message sending out its residual energy and the current position. After the level-1 CHs (denoted as $n_{CHj-1}$) receive this message, they can construct their active node-set, denoted as $S_1(t)$, and then will broadcast a "command" message with total residual energy of the cluster and the total reciprocal distance within the cluster of each node from the cluster head and also cardinality of $S_1(t)$ denoted as $N_1$. Each node will then set its probability of becoming a level-2 cluster-head according to the formula,

$$p_2(u) = \sqrt{N_1}\left[\phi\frac{E_u(t)}{\sum_{v\in S_1(t)}E_v(t)} + (1-\phi)\left(\frac{1/dis(n_u, n_{CH1})}{\sum_{v\in S_1(t)}1/dis(n_v, n_{CH1})}\right)\right]$$

Based on this probability, nodes elect whether they will become level-2 CHs or not.

(vi) Generalizing, the clustering may extend up to level-j. For any level-(j-1) CH, the CH would broadcast a level-(j-1) CH message with the radio range $\frac{R}{\sqrt{N_{CH1}^{opt}\prod_{j=1}^{j=2}\sqrt{N_j(t)}}}$.

Each regular node that receives this message will send its current position as well as the energy level to CH with a "join" message. After the level-(j-1) CHs (denoted as $n_{CHj-1}$) receive this message, they can construct their active node-set, denoted as $S_{j-1}(t)$, and then will broadcast a "command" message with total residual energy of the cluster and the total reciprocal distance within the cluster of each node from the cluster-head and also cardinality of $S_{j-1}(t)$ denoted as $N_{j-1}$. Each node will then set its probability of becoming a level-j cluster-head according to the formula,

$$p_j(u) = \sqrt{N_{j-1}}\left[\phi\frac{E_u(t)}{\sum_{v\in S_{j-1}(t)}E_v(t)} + (1-\phi)\left(\frac{1/dis(n_u, n_{CHj-1})}{\sum_{v\in S_{j-1}(t)}1/dis(n_v, n_{CHj-1})}\right)\right]$$

The process of clustering will stop if a node has two or less number of nodes in its regular set.

(vii) In addition to this, each regular node also maintains information about the closest CH that has been heard from. The closest CH may be from the higher levels. Whenever a regular node receives a level-i CH message, it checks whether the sender is closer than the closest CH heard from until current time. If it is, the closest CH information is set to current CH. In any case, the regular node will send out a join message to the present CH. However, the final decision to join a CH is deferred until finalization of topology.

(viii) After all CHs have been chosen, each regular node checks whether it has joined the closest CH (which may be from higher levels). If not, it sends a message to the new CH indicating that it will now join that cluster and sets its level to one more than the level of CH. It also sends a de-join message to previous CH.

**(b) Network Operation Phase**

After level-T clustering topology is formed, the regular nodes start transmitting the sensed data to their CHs. Level-T CHs aggregate the sensed data and send it to level-(T-1) CHs and so forth. Finally, all level-1 CH transmit the aggregated data to sink node. The cost of delivering data to the sink is sum of energy spent by nodes to send the data to their respective CH.

An obvious advantage in LAMC is the decrease in the average number of hops required to transmit sensed data to the base station in a network. By allowing regular nodes to join higher level CHs, the depth of topology can be decreased. Also, since the regular nodes are now transmitting to the closest CHs, the energy consumption of these nodes will also decrease.

### 4.2 PAMC

LAMC presented above, does not do away with the location related assumption. For this we present a new algorithm Power Aware Multi-level Clustering (PAMC). Here we make additional assumption that the nodes have many discrete power levels that determine their transmission range. An example of such nodes is the Berkley motes [9]. The assumption about power levels is derived from [7] where the authors have used the concept of Minimum Reachability Power. Let $MinPwr_i$ be the minimum power level required by the node i to reach its cluster-head. The TRMRP (Total Reciprocal Minimum Reachability Power) is $TRMRP = \sum 1/MinPwr_i$ . The MRP of any node can give us an idea about the distance of transmission and can replace the distance parameter in the formula of LAMC. Note that MRP is used to decrease the intra-cluster communication cost within the hierarchy. The protocol assumes that for any Power Level $L_i$, there is a corresponding Transmission Range $R_i$ such that

$$R_i < R_j \quad \forall \quad L_i < L_j$$

The operation of the protocol is divided into two phases.

**(a) Cluster Setup Phase**

(i) At the beginning of each round, all nodes are set to regular (non-CH status). The cluster setup phase is initiated by the base station which sends out a Start beacon.

(ii) On receiving this Start beacon, each nodes selects its minimum power level needed to reach the base station and stores it in the memory for subsequent rounds. The discovery of the power level is done by recursively sending signals of lower power levels till acknowledgement is not received from base station. Note that the process of MRP discovery to base station is done only once during entire network lifetime and is cached for subsequent usage.

(iii) The active nodes receiving the Start message reply by sending out their residual energy $E_u(t)$ and the MRP $P_u$ to the sink node to indicate that level-1 CH will now be selected. $P_u$ can be thought as a measure of distance between source and sink.

(iv) After the sink node receives these values, it will broadcast a ''command'' message with the following two values: one is the total remaining energy of the network $\sum E_v(t)$ and the other is $\sum 1/P_v$ the TRMRP from all active nodes.

(v) On receiving this ''command'' message, each active node u sets its probability of becoming a level-1 CH, denoted as $P_1(u)$, by using the following formula,

$$p_1(u) = N_{CH1}^{opt}\left[\phi \frac{E_u(t)}{\sum_{v\in S(t)} E_v(t)} + (1-\phi)\left(\frac{1/P_u)}{\sum_{v\in S(t)} 1/P_v)}\right)\right]$$

Here, $\phi$ is a parameter which defines how much weightage is given to residual energy and how much to distance of node from sink and $N_{CH1}^{opt}$ is the optimal number of cluster heads for level-1. Thus, nodes with higher energy or less distance will have a greater probability of becoming the cluster heads. Nodes thus elect whether to become a level-1 cluster head or not based on global knowledge.

(vi) After this, in order to reduce energy consumption, all level-1 CH send out a level-1 CH message with the Power level equivalent to range $\frac{R}{\sqrt{N_{CH1}^{opt}}}$ where R is the Range associated with MRP. Any node receiving this message replies with a "join" message sending out its residual energy and MRP to the CH. Note that the regular nodes can cache some of the entries of MRP to other nodes to reduce the message overhead in calculating the MRP

(v) The regular nodes keep track of the closest CH from which they have heard till now. The closest CH may be from higher levels. Closeness is now defined in terms of MRP. If two nodes have same MRP, the node that was heard earlier from (i.e. the node higher up in the hierarchy) is given preference.

(vi) After the level-1 CHs (denoted as $n_{CH1}$ receive this message, they can construct their active node-set, denoted as $S_i(t)$, and then will broadcast a "command" message with total residual energy of the cluster, the TRMRP within the cluster of each node from the cluster-head and the cardinality of its active set $N_1$. Each node will then set its probability of becoming a level-2 cluster-head according to the formula,

$$p_2(u) = \sqrt{N_1}\left[\phi \frac{E_u(t)}{\sum_{v \in S_1(t)} E_v(t)} + (1-\phi)\left(\frac{1/P_u}{\sum_{v \in S_1(t)} 1/P_v}\right)\right]$$

Based on this probability, nodes elect whether they will become cluster heads for level-2 or not.

(vi) Generalizing, the clustering may extend up to level-j. For any level-(j-1) CH, the CH would broadcast a level-(j-1) CH message with the power-level equivalent to radio range $\frac{R}{\sqrt{N_{CH1}^{opt} \prod_{j=1}^{j=2} \sqrt{N_j(t)}}}$. Each regular node that receives this message will send its MRP as well as the energy level to CH with a "join" message. After the level-(j-1) CHs (denoted as $n_{CHj-1}$) receive this message, they can construct their active node-set, denoted as $S_{j-1}(t)$, and then will broadcast a "command" message with total residual energy of the cluster and the TRMRP within the cluster of each node from the cluster-head and also cardinality of $S_{j-1}(t)$ denoted as $N_{j-1}$. Each node will then set its probability of becoming a level-j cluster-head according to the formula,

$$p_j(u) = \sqrt{N_{j-1}}\left[\phi \frac{E_u(t)}{\sum_{v \in S_{j-1}(t)} E_v(t)} + (1-\phi)\left(\frac{1/P_u}{\sum_{v \in S_{j-1}(t)} 1/P_v}\right)\right]$$

The process of clustering will stop if a node has two or less number of nodes in its regular set.

(viii) Once the clustering process is over, the regular nodes select their final CHs based on the closeness criteria explained before. If a regular node needs to change its CH, it will send out control messages to both old and new CH to indicate its change of preferences.

(b) **Network Operation Phase**

After level-T clustering topology is formed, the regular nodes start transmitting the sensed data to their CHs. Level-T CHs aggregate the sensed data and send it to level-(T-1) CHs and so forth. Finally, all level-1 CH transmit the aggregated data to sink node.

The cost of delivering data to the sink is sum of energy spent by nodes to send the data to their respective CH. It may be noted that the process of storing some of the last known MRPs can greatly reduce the communication cost of the network.

## 5 SIMULATION AND RESULTS

### 5.1 Simulation Environment

We have simulated EEMC, LAMC and PAMC using Qualnet, a discrete event based object oriented simulator. Table I lists the simulation parameters used. The very small initial energy of 0.1J is chosen to shorten the simulation time. To evaluate the algorithms, we have used the following performance metrics:

First Node Dead and Half Nodes Dead – These are a number of rounds that elapse before first node and half of the nodes, respectively, run out of energy. It gives us an idea about the network lifetime, since the network will be operational only until more than half of the nodes are sensing the data.

Overhead Ratio – The ratio of control packets to the data packets transmitted per round is an indication of the control overhead incurred by the algorithm. Lower value of this ratio indicates better algorithm.

Latency per Packet – The average number of hops in the network is used as a measure to evaluate the latency of the algorithm. A larger value of average number of hops indicates greater delay in packet reception by the base station.

In addition to this, we will also study the impact of parameter $\phi$ on the network lifetime of both PAMC and LAMC, and also impact of different cache sizes on the performance of PAMC algorithm.

### 5.2 Simulation Results

Many runs of each algorithm are simulated on different random distribution of nodes and the results are recorded as average of all the runs. A more efficient scheme is denoted by greater network lifetime and less number of hops.

TABLE I
SIMULATION PARAMETERS

| | |
|---|---|
| Number of nodes | 100 - 500 |
| Network Grid | (0, 0) to (1000, 1000) |
| Base Station Position | (500, 500) |
| Energy Per Node | 0.1J |
| Size of Data Packet | 500 bits |
| Size of Control Packet | 10 bits |
| Number of cached entries (for PAMC) | 10 |

**Lifetime Analysis**

It is clear from Fig. 1 and Fig. 2 that the LAMC outperforms both the EEMC and the PAMC in terms of network lifetime. An obvious reason for this is that LAMC is derived from EEMC and optimizes the energy usage of EEMC by choosing the closest CH rather than the latest CH.

The PAMC is found to decrease the network life-time slightly, since it has to incur an overhead in terms of MRP calculations by repeated message sending. The performance of EEMC and PAMC are, however, comparable since in EEMC we have not considered the energy overhead in obtaining the current location information via GPS.

Fig. 2 shows that PAMC outperforms EEMC for 100 nodes because the use of cache (size ten) decreases the MRP queries. As the number of nodes increases, the need to query for MRP also increases. This increases the energy overhead.

By doing away with the GPS, it is possible to miniaturize the nodes to an even greater extent.

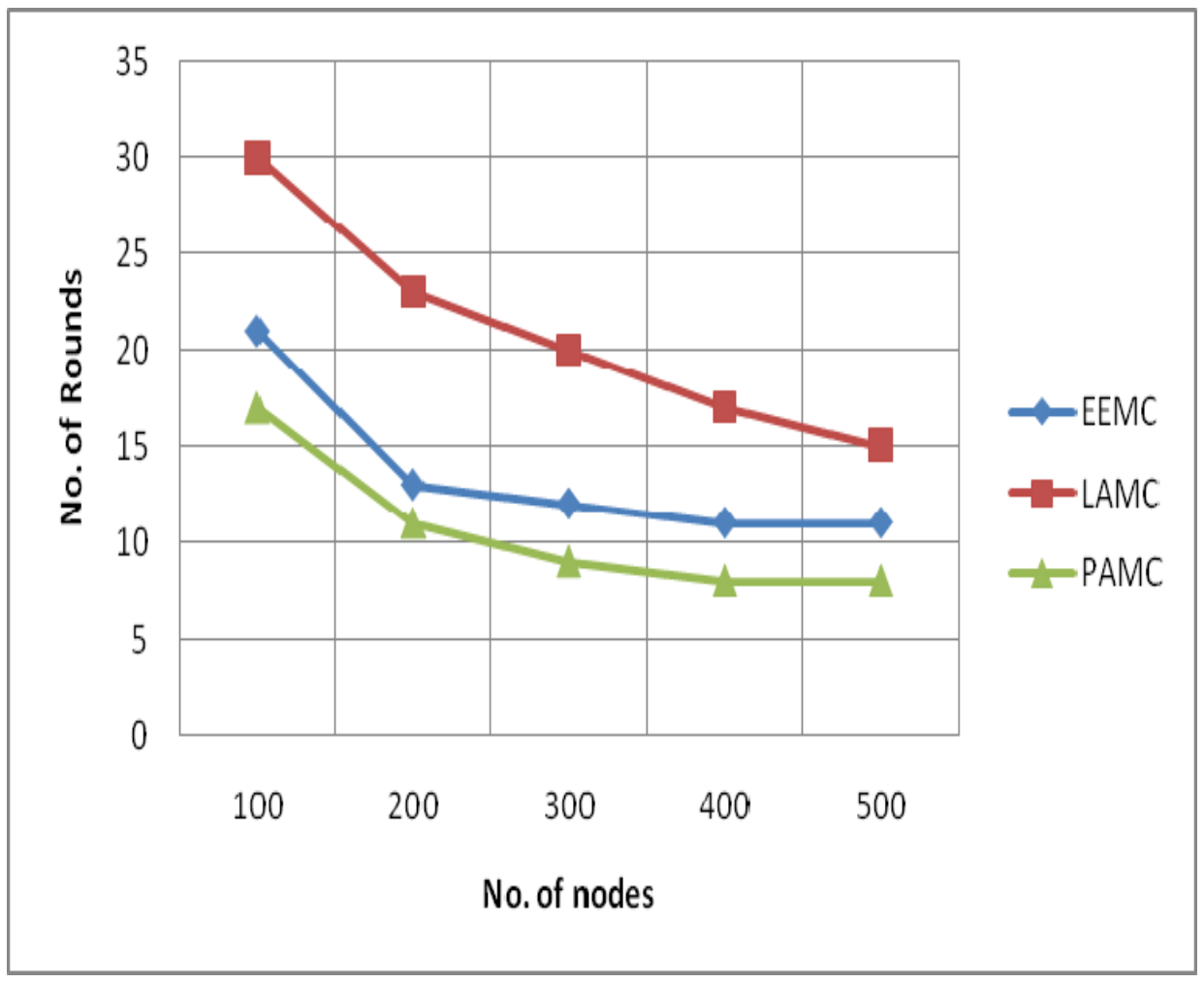

Fig. 1. The FND analysis.

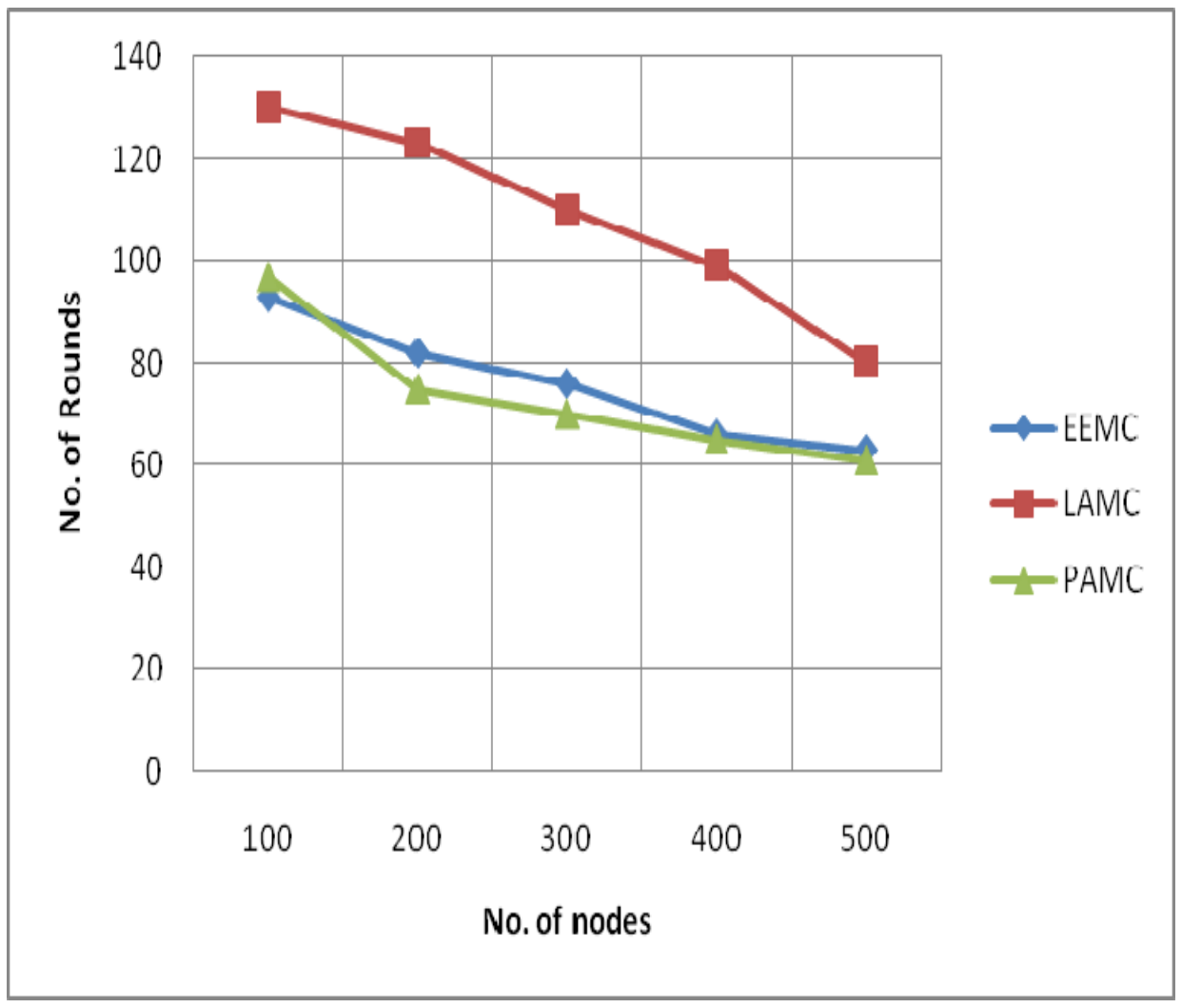

Fig. 2. The HND analysis.

**Overhead analysis**

The histogram in Fig. 3 clearly shows that EEMC has the minimum control overhead over all the algorithms. This is understandable since LAMC needs to send extra control messages to make the final selection of CH. Recall that in LAMC the regular nodes select their final CH after the cluster-setup phase has ended. Control packets need to be transmitted to inform the old and new CHs about the change.

PAMC has even higher overhead ratio since the query for MRP entails sending of more beacons to the prospective CHs. However, for 100 nodes the PAMC shows lesser ratio since the use of cache limits the transmission of MRP queries. This result is in line with the HND results which show PAMC to be better than EEMC for 100 nodes.

**Latency analysis**

Fig. 4 clearly shows that LAMC and PAMC both decrease the latency of the network by optimizing the cluster formation. Since both these algorithms allow regular-nodes to join higher level CHs, it leads to a more compact topology which is reflected with a decrease in average number of hops.

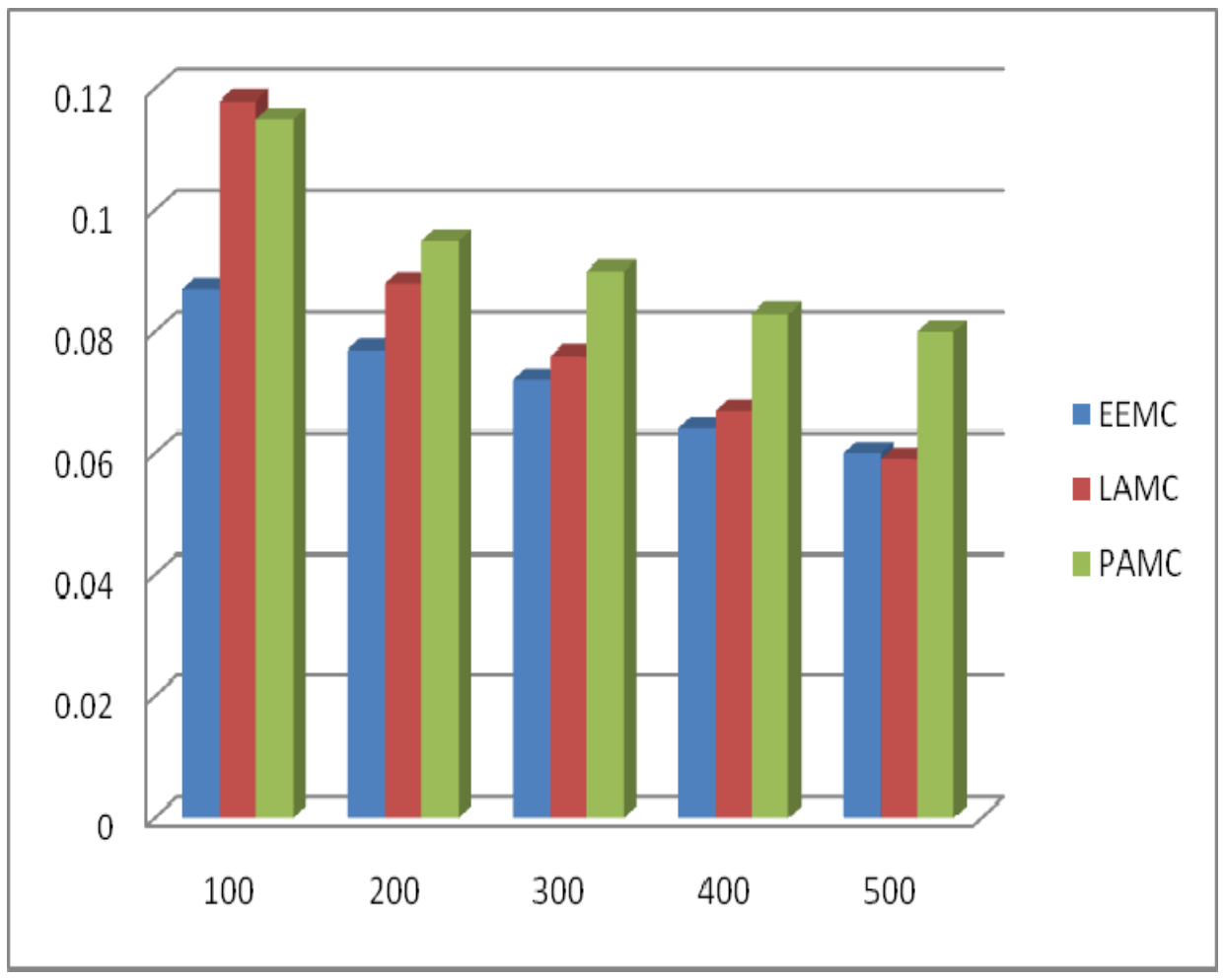

Fig. 3. Control-to-data ratio analysis.

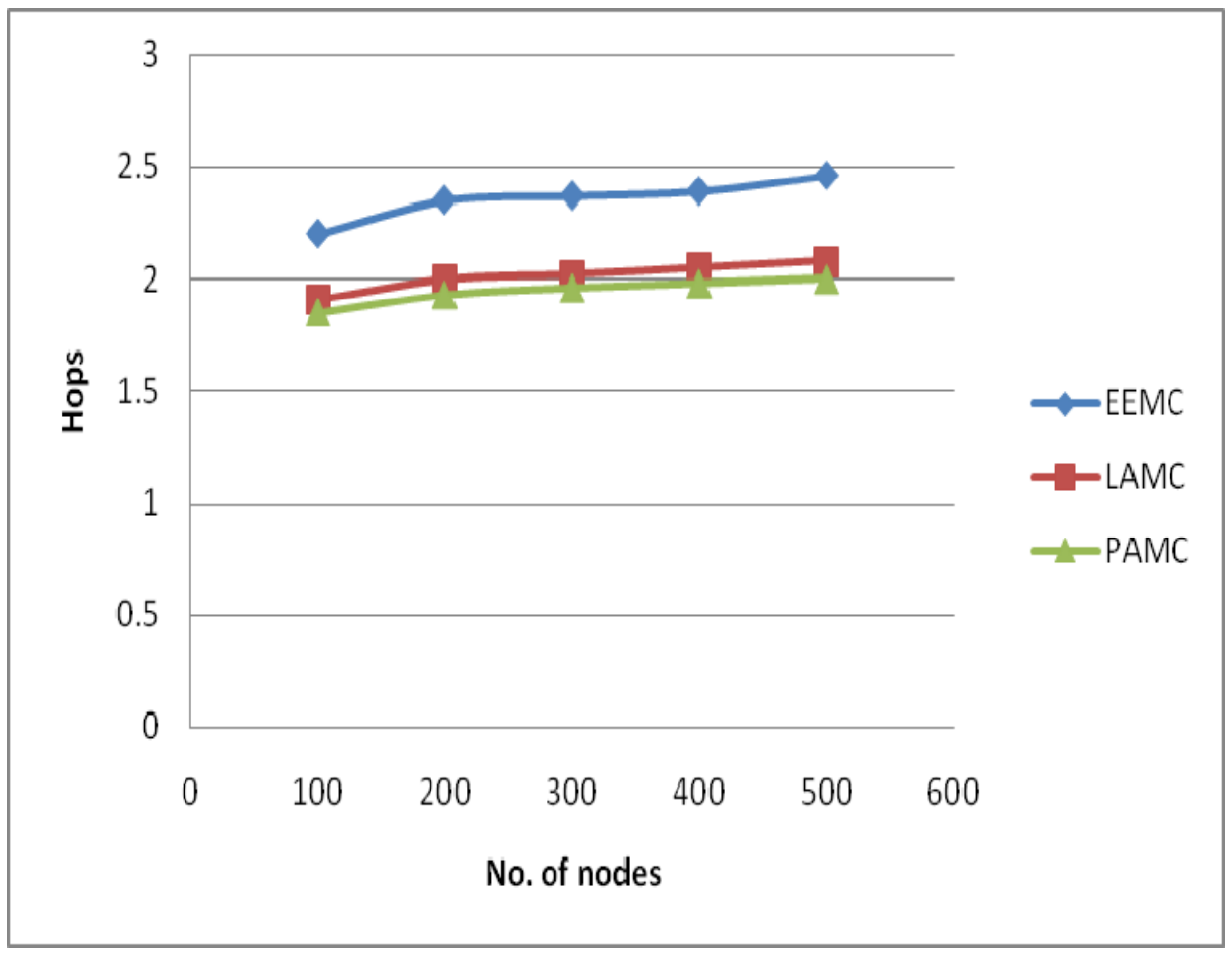

Fig. 4. Average number of hops.

PAMC leads to a further decrease in the number of hops since the criteria of "closeness" is now MRP. For any regular nodes, all the CHs will lie in one of the six power levels. Hence there will be frequent ties in which case higher CHs will be preferred. By contrast, LAMC will have discrete values of distance which will lead to fewer ties and hence the closer but lower CHs may be selected.

**Impact of parameter $\phi$**

To study the impact of parameter $\phi$ we simulate the algorithms for 100 nodes. As is clear from the graph in Fig. 5, for LAMC, the value of $\phi$ plays an important role in determining the network lifetime. However, the impact is highly irregular. The general trend is that lower the value of $\phi$, lower is the network lifetime. The network lifetime peaks when $\phi$ reaches 0.8 and then decreases again. Note that by lifetime we mean the numbers of rounds that have elapsed before half of the nodes are dead.

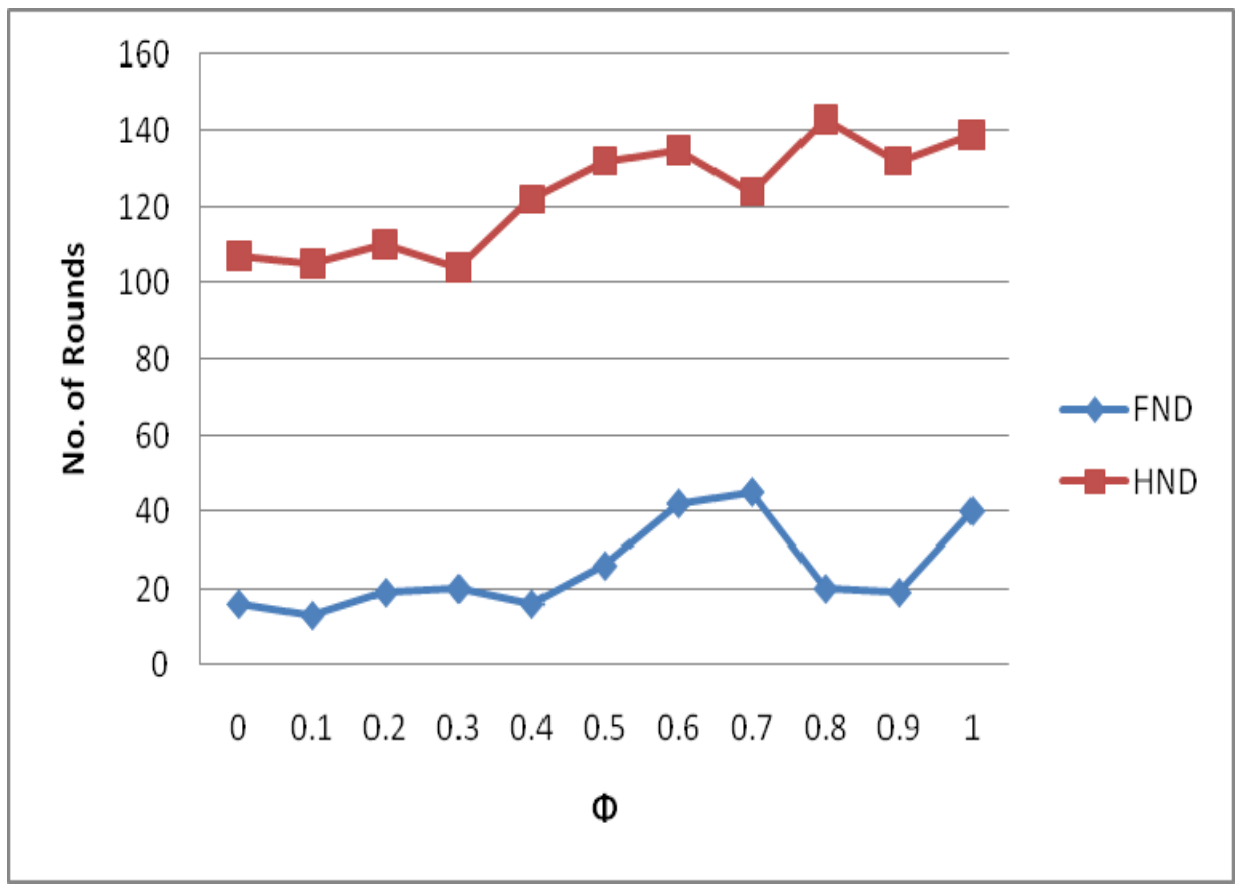

Fig. 5. Impact of $\phi$ on LAMC.

For PAMC, the trend is the same. There is a general increase in lifetime as $\phi$ increases. The maximum lifetime is obtained in this case for 0.7 and there is a dip in the network lifetime for 0.8.

**Impact of caching on PAMC**

As already discussed, in PAMC algorithm each node can store some of the latest MRP values in the cache. If the cache size is bigger, lesser MRP queries will have to be sent out to obtain MRP information. This will decrease the control overhead and consequently increase the network lifetime. However, since WSN are memory constrained, there is little scope for increasing cache size beyond a few bytes.

The histogram shown below corroborates this view. As we increase the cache size, there is a small increase in the HND value, which signifies an increase in the network lifetime. Values have been obtained for cache size of 1, 5, 10 and 20 entries.

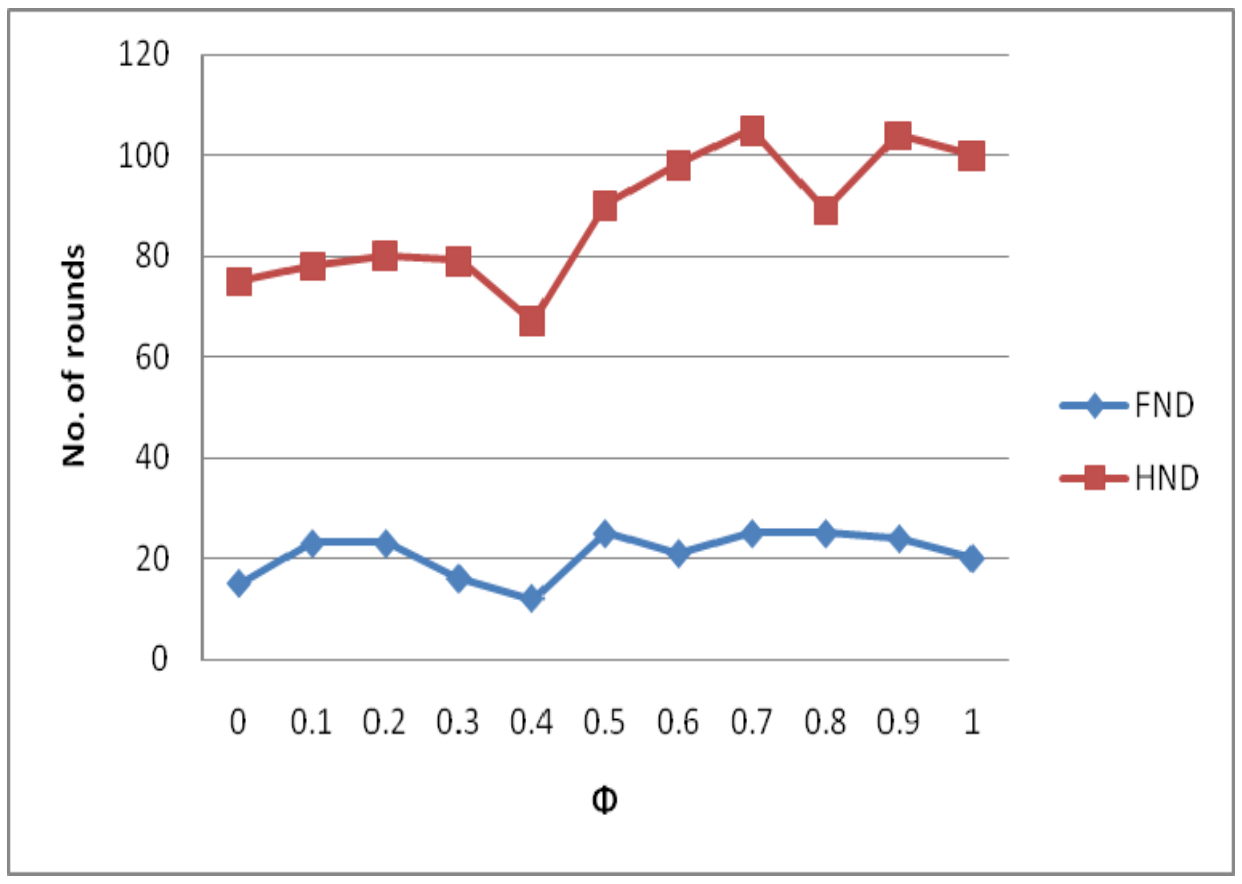

Fig. 6. Impact of $\phi$ on PAMC.

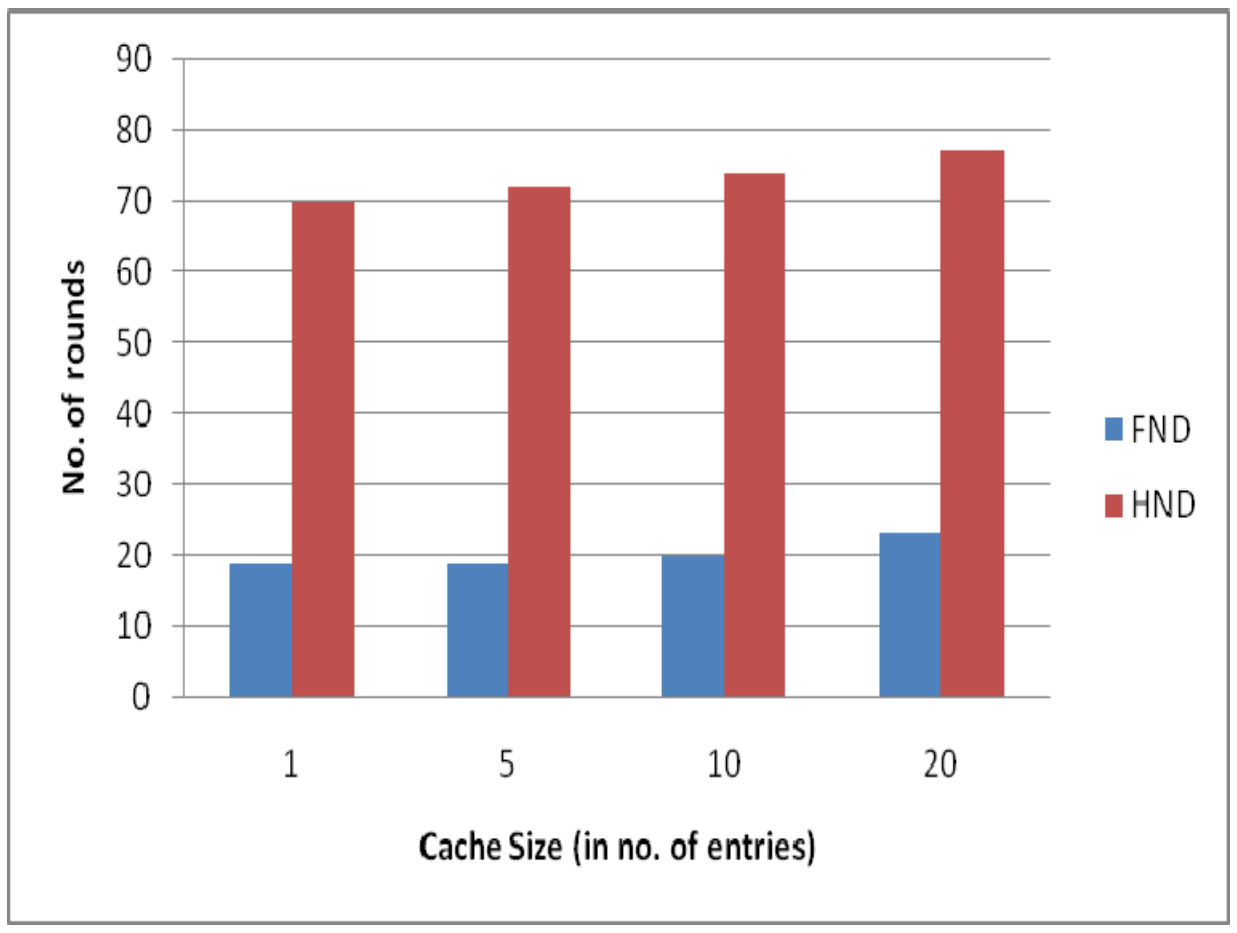

Fig. 7. Impact of cache on PAMC.

## 6 CONCLUSIONS

This paper presents two multi-level clustering algorithms namely LAMC and PAMC for wireless sensor networks. Both these algorithms are built upon EEMC algorithm proposed in [5] and aim to further prolong the lifetime of WSN by minimizing the energy consumption of the network.

While LAMC reduces the latency of the network, the PAMC removes the constraint of location awareness altogether.

Simulations have been performed to study the performance of these algorithms. Results show that LAMC is more efficient than EEMC whereas PAMC gives comparable performance without the need of GPS fitting at each node.

The PAMC algorithm that we have discussed can be optimized further by making use of alternative MRP evaluation techniques. In addition, to remove location awareness criteria of LAMC, localization techniques can be used that would require only some of the nodes to be equipped with GPS.

**Surender Soni** received his BTech Degree from NIT Hamirpur, and MTech Degree from Punjab University Chandigarh. He is currently a Research Scholar at the Department of Computer Science & Engineering, NIT Hamirpur. His research interests include resource management in wireless sensor networks.

**Dr. Narottam Chand** received his PhD degree from IIT Roorkee in Computer Science and Engineering. Previously he received MTech and BTech degrees in Computer Science and Engineering from IIT Delhi and NIT Hamirpur respectively.

Presently he is working as Head, Department of Computer Science and Engineering, NIT Hamirpur. He also served as Head, Institute Computer Centre, NIT Hamirpur from February 2008 to July 2009. He has coordinated different key assignments at NIT Hamirpur like Campus Wide Networking, Institute Web Site, Institute Office Automation.

His current research areas of interest include mobile computing, mobile ad hoc networks and wireless sensor networks. He has published more than 50 research papers in International/National journals & conferences and guiding six PhDs in these areas. He is member of ISTE, CSI, International Association of Engineers and Internet Society.